\documentclass[superscriptaddress,aps,preprintnumbers,amsmath,showpacs,amssymb,prd,nofootinbib,reprint,twocolumn,floatfix]{revtex4-1}

\usepackage{graphicx}
\usepackage[colorlinks,allcolors=blue]{hyperref} 
\usepackage{amsmath,amssymb,bm}
\usepackage[T1]{fontenc}
\usepackage{mathtools}

\begin{document}
\title{Nearly Monochromatic Primordial Black Holes as total Dark Matter from Bubble Collapse}

\author{Haonan Wang}
\email{haonan\_cosmos@tongji.edu.cn}
\affiliation{School of Physics Science and Engineering, Tongji University, Shanghai 200092, China}
\author{Ying-li Zhang}
\email{yingli@tongji.edu.cn}
\affiliation{School of Physics Science and Engineering, Tongji University, Shanghai 200092, China}
\affiliation{Institute for Advanced Study of Tongji University, Shanghai 200092, China}
\affiliation{Kavli Institute for the Physics and Mathematics of the Universe (WPI), Chiba 277-8583, Japan}
\affiliation{Center for Gravitation and Cosmology, Yangzhou University, Yangzhou 225009, China}
\author{Teruaki Suyama}
\email{suyama@phys.sci.isct.ac.jp}
\affiliation{Department of Physics, Institute of Science Tokyo, 2-12-1 Ookayama, Meguro-ku, Tokyo 152-8551, Japan}

\begin{abstract}
We propose a two-field model where the inflaton $\chi$ is non-minimally coupled to the instanton $\phi$. By choosing an appropriate coupling function, we realize the scenario where the difference of the values of potential between false vacuum (FV) and true vacuum (TV) is maximized during inflation. Most of the bubbles are created at this time. After inflation ends, the potential value of FV drops below that of TV so that these bubbles collapse to form primordial black holes (PBHs). By tuning the parameters of our model, we analyze the Coleman-de Luccia (CDL) and Hawking-Moss (HM) process, finding that the corresponding mass function of PBHs is sharply peaked, implying that we can realize either PBHs as cold dark matter, sub-solar PBHs, or
supermassive PBHs in this scenario without enhancement of primordial curvature perturbations. 

\end{abstract}

\maketitle

\section{Intruduction}
\label{sec:intro}
Primordial black holes (PBHs) are black holes that may form in the early universe~\cite{Zeldovich:1967lct,Hawking:1971ei,Carr:1974nx}. Their potential cosmological implications have attracted much attention recently. Since PBHs interact only via gravity, they serve as natural candidate for cold dark matter (CDM) without requiring physics beyond the Standard Model~\cite{Chapline:1975ojl,Niikura:2019kqi,Katz:2018zrn,Smyth:2019whb}.
They also provide a possible explanation for the origin of the mergers of black hole (BH) binaries detected by LIGO/Virgo/KAGRA gravitational wave (GW) observations~\cite{Bird:2016dcv,Sasaki:2016jop,Clesse:2016vqa,Kimura:2021sqz,Wang:2022nml,Franciolini:2022tfm,Franciolini:2023opt,LIGOScientific:2018mvr,LIGOScientific:2020ibl,KAGRA:2021vkt}, and the seeds of supermassive BHs. Moreover, evaporation of ultra-light PBHs with mass lighter than $5\times10^8$g can realize the reheating and early matter-dominated epochs with detectable signatures in the stochastic GW background~\cite{Hidalgo:2011fj,Martin:2019nuw,Papanikolaou:2020qtd,Domenech:2020ssp,RiajulHaque:2023cqe,Domenech:2024wao,del-Corral:2025fca,Wang:2025lti}.


Motivated by these physical phenomenology, there have been a lot of proposals on the mechanisms of PBH formation~\cite{Khlopov:2008qy,Carr:2016drx,Sasaki:2018dmp}. One typical mechanism requires the enhancement of primordial curvature perturbations on small scales~\cite{Zeldovich:1967lct,Hawking:1971ei,Carr:1974nx,Polnarev:1985btg,Yokoyama:1995ex,Garcia-Bellido:1996mdl,Kawasaki:1997ju,Yokoyama:1998pt,Pi:2017gih,Tada:2019amh,Pi:2021dft,Pi:2022zxs,Wang:2024vfv,Kim:2025dyi,Wang:2025dbj,Cacciapaglia:2025xqd,Carr_2020,Khlopov_2010,Sasaki_2018,Carr_2021,Green_2021,Villanueva_Domingo_2021,Escriv__2024,Zeng:2025law}. A common feature of these models is the significant enhancement of the amplitude of GWs at the second order induced by the enhanced curvature perturbations, which may serve as a candidate for the detection of stochastic GW background by pulsar timing array (PTA) collaborations~\cite{Saito:2008jc,Cai:2018dig,Domenech:2020ers,Balaji:2023ehk,NANOGrav:2023hvm,Reardon:2023gzh,EPTA:2023fyk,Xu:2023wog,Lozanov:2023rcd,Domenech:2024rks,Inui:2024fgk,Dong:2015yjs}.


Alternatively, PBHs can arise from non-perturbative quantum effects in the early universe, including the collapse of cosmic strings~\cite{Garriga:1992nm,Caldwell:1995fu}, domain walls~\cite{Khlopov:2008qy,Deng:2016vzb,Liu:2019lul}, false vacuum relics \cite{Cai:2024nln,Maeso:2021xvl}, and the collisions of vacuum bubbles~\cite{Hawking:1982ga,Sato:1981gv,Kodama:1982sf,Crawford:1982yz,Blau:1986cw,Khlopov:1998nm,Johnson:2011wt,Yuwen:2024gcf,Liu:2021svg,Bian:2025twi}. In particular, PBH formation via collapse of vacuum bubbles nucleated during inflation was proposed in~\cite{Garriga:2015fdk}. Assuming a constant tunneling rate, numerical studies found that the resulting PBH mass function in the radiation-dominated era is rather broad so that they cannot account for all DM~\cite{Deng:2017uwc,Ashoorioon:2020hln}. Later study~\cite{Kusenko:2020pcg} extended this framework to scenarios where inflation transitions into an intermediate matter-dominated phase, leading to a multi-step PBH mass function to relieve this problem. However, in either case, it is difficult to realize a nearly monochromatic mass spectrum of PBHs.

In this work, we adopt the vacuum bubble collapse scenario~\cite{Garriga:2015fdk,Deng:2017uwc} but introduce a time-dependent tunneling rate that maximizes during inflation. This is achieved via a two-field model where an instanton field $\phi$ couples non-minimally to the inflaton $\chi$ through a  function $g(\chi)$. This function reaches the maximum value during inflation, at which the difference in the potential value between false vacuum (FV) and true vacuum (TV) maximizes. Consequently, the PBH mass function exhibits a sharp peak, allowing for a nearly monochromatic PBH spectrum. By tuning the parameters, we can realize either PBHs as all DM, stellar-mass BHs, or supermassive BHs.


\section{PBHs formed by collapsing bubbles}
\label{sec:pbhFromBubble}
In this section, we briefly review the scenario where PBHs form from the collapse of bubbles and initialize the motivation of the corresponding two-field model. We consider an FLRW metric outside the bubble
\begin{equation}
    {\rm d}s^{2}=-{\rm d}t^{2}+a^{2}(t)({\rm d}r^{2}+r^{2}{\rm d}\Omega^{2})\,,
\end{equation}
where the Hubble parameter $H_{\rm inf}$ is nearly a constant and the scale factor $a\propto\exp(H_{\rm inf}t)$ during the inflationary epoch $t_i<t<t_f$. After a bubble is nucleated at time $t_{\rm n}$, the world line of the bubble wall can be approximately described by a null curve so that its physical size at time $t>t_{\rm n}$ during inflation is estimated by
\begin{equation}\label{eq:R(t)}
    R(t,t_{\rm n})=\frac{1}{H_{\rm inf}}\left[e^{H_{\rm inf}(t-t_{\rm n})}-1\right]\approx \frac{1}{H_{\rm inf}}e^{H_{\rm inf}(t-t_{\rm n})}\,.
\end{equation}
This implies that after nucleation, the bubble radius rapidly increases and exceeds the Hubble horizon scale $H_{\rm inf}^{-1}$ after approximately one e-fold expansion. 


Meanwhile, since the abundance of these bubbles is severely diluted by the inflating scale factor, the number density of bubbles whose physical radii are within the range $(R, R+{\rm d}R)$  at time $t$ is
\begin{equation}
    \frac{{\rm d}n(R,t)}{{\rm d}R}=\eta(t_{\rm n})\Gamma(t_{\rm n})\frac{a^{3}(t_{\rm n})}{a^{3}(t)}\left | \frac{{\rm d}t_{\rm n}}{{\rm d}R} \right |\,,
\end{equation}
where $t_{\rm n}$ can be expressed by $R$ via Eq.~\eqref{eq:R(t)},  $\Gamma(t)$ is the bubble nucleation rate at time $t$, $\eta(t_{\rm n})$ is the FV fraction at time $t_{\rm n}$ given by
\begin{equation}
    \eta(t_{\rm n})=\exp\left[-\frac{4\pi}{3}\int_{t_i}^{t_{\rm n}}{\rm d}t\,\frac{\Gamma(t)}{H_{\rm inf}^3}\right]\,.
\end{equation}
For simplicity, we do not consider any collision between bubbles. This implies that the nucleation rate per unit Hubble volume $\Gamma/H_{\rm inf}^{4}\ll1$, implying $\eta(t_{\rm n})\approx1$. Hence, at the end of inflation, the number density of bubbles is expressed as
\begin{equation}
    \frac{{\rm d}n}{{\rm d}R_f}=\frac{\Gamma(t_{\rm n})}{H_{\rm inf}^{4}}\frac{1}{R_f^{4}}\,,
    \qquad
    R_f\equiv R(t_f)\,.
\end{equation}

After inflation, those bubbles may collapse to form PBHs. For instance, Ref.~\cite{Garriga:2015fdk} presumes that the FV energy outside the bubbles turns into radiation immediately. Then the expanding bubbles gradually transform their kinetic energy to the surrounding radiation and collapse due to the wall tension and internal negative vacuum pressure. For ``supercritical'' bubble whose size exceeds the Hubble horizon $H_{\rm inf}^{-1}$ during inflation%
\footnote{Here we ignore the ``subcritical'' case since we are mainly interested in the case where the mass of PBHs $M\gtrsim10^{17}$g. Moreover, a small subcritical bubble may be deformed due to quantum fluctuations \cite{Garriga:1991tb}, which prevents  black hole formation and yields a lower bound $M_{\rm min}$ for the resulting PBH mass range.}%
, the upper bound of the resulting PBH mass $M$ can be determined by demanding that the Schwarzschild radius of the BH cannot exceed the radius of the comoving FRW region affected by the bubble wall when it comes within the cosmological horizon~\cite{Deng:2016vzb}, which roughly yields
\begin{equation}\label{eq:M(R)}
    M\sim M_{\rm Pl}^2H_{\rm inf}R_f^2\,.
\end{equation}
Further numerical simulation in~\cite{Deng:2017uwc} has proved \eqref{eq:M(R)}.

Besides the mechanism above, if the energy density outside the bubble drops down to a value lower than that inside~\cite{Kusenko:2020pcg,Ai:2024cka,Murai:2025hse}, it will also stop the expansion of bubbles and force them to collapse, since the pressure on the bubble wall is always along the direction from the region with lower energy density to higher one.


For definiteness, we adopt \eqref{eq:M(R)} to calculate the PBH mass. The fraction of PBH is defined in terms of PBH mass function as
\begin{equation}\label{fPBH}
    f_{\mathrm{PBH}}\equiv\frac{\rho_{\mathrm{PBH}}(t)}{\rho_{\mathrm{CDM}}(t)}=\int\frac{{\rm d}M}{M}f(M)\,,
\end{equation}
where $\rho_{\mathrm{CDM}}(t)$ and and $\rho_{\mathrm{PBH}}(t)$ are the mass density of CDM and PBH, respectively. The mass function reads
\begin{equation}\label{eq:f(M)}
    f(M)=\frac{M^2}{\rho_{\mathrm{CDM}}(t)}\frac{{\rm d}n(t)}{{\rm d}M}
    \sim \frac{\Gamma(t_{\rm n})}{{H}_{\rm inf}^{4}} C\left(\frac{\mathcal{M}_{\mathrm{eq}}}{M}\right)^{\frac{1}{2}}\,,
\end{equation}
where we have used
\begin{equation}
    \rho_{\mathrm{CDM}}(t)\sim\frac{M_{\rm Pl}^3}{Ct^{3/2}\mathcal{M}_{\mathrm{eq}}^{1/2}}\,,
\end{equation}
with $C\sim 10$ a constant and $\mathcal{M}_{\mathrm{eq}}\sim10^{17}~M_{\odot}$ as the total mass inside the Hubble radius at radiation-matter equality time $t_{\mathrm{eq}}$. Using \eqref{eq:R(t)} and \eqref{eq:M(R)}, $\Gamma(t_{\rm n})$ can be expressed as a function of PBH mass $M$. It is evident from \eqref{eq:f(M)} that if the bubble nucleation rate $\Gamma$ is constant, the PBH mass spectrum $f(M)$ will be smooth and expand a broad mass range. Hence, it is seriously constrained by observations and cannot contribute to most of CDM. 

However, if $\Gamma$ varies in time significantly during inflation to dominate over the suppression factor $M^{-1/2}$ in \eqref{eq:f(M)}, the above situation may be changed so that a sharp peak appears in $f(M)$ to realize the scenario where PBHs perform as the main candidate of CDM. In the following, we construct a concrete model to realize this scenario.


\section{The model}
\label{sec:model}

We consider a two-field model as follows
\begin{eqnarray}
    \mathcal{L}=-\frac{1}{2}\partial _{\mu}\chi\partial^{\mu}\chi-\frac{1}{2}\partial _{\mu}\phi\partial^{\mu}\phi-V(\chi,\phi)\,.
\end{eqnarray}
The potential $V(\chi,\phi)=V_{\chi}(\chi)+V_{\phi}(\chi,\phi)$ with
\begin{align}
    V_{\chi}(\chi)&=\frac{1}{2}m^2\chi^2\,,\label{eq:V}\\
    V_{\phi}(\chi,\phi)&=\lambda{(\phi^{2}-v^{2})}^{2}+g(\chi) v^2(\phi-v)^{2}\,,\label{eq:Vphi}
\end{align}
where $m$, $\lambda$ and $v$ are three parameters, $g(\chi)$ is a function of scalar field $\chi$. We let $\chi$ and $\phi$ be two scalar fields which play the roles of the inflaton and instanton field, respectively. 
\begin{figure}[ht]
    \includegraphics[width=0.95\linewidth]{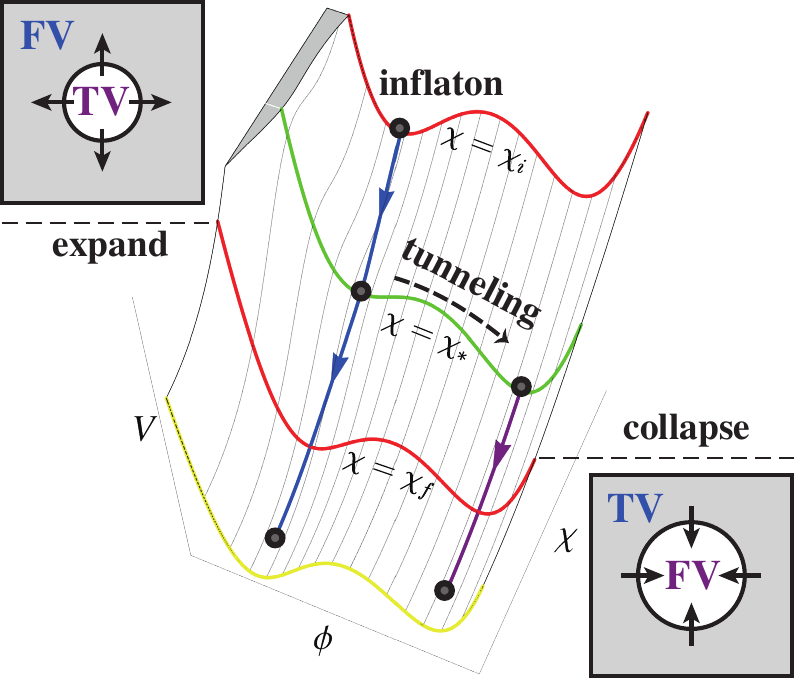}
    \caption{Schematic illustration of the two dimensional potential $V(\chi,\phi)$. An inflaton starting from $\chi=\chi_i$ can either roll down along the blue trajectory all the time, or tunnel through the barrier and then roll down along the purple trajectory, resulting in bubble nucleation.
    \label{fig:mechanism}}
\end{figure}

During inflation, the trajectory in the field space can be decomposed into $\chi$ and $\phi$ directions.
Principally speaking, the tunneling process along $\phi$ direction happens at each point when the inflaton $\chi$ rolls down, as shown in Fig.~\ref{fig:mechanism}.

\begin{figure}[ht]
    \includegraphics[width=0.85\linewidth]{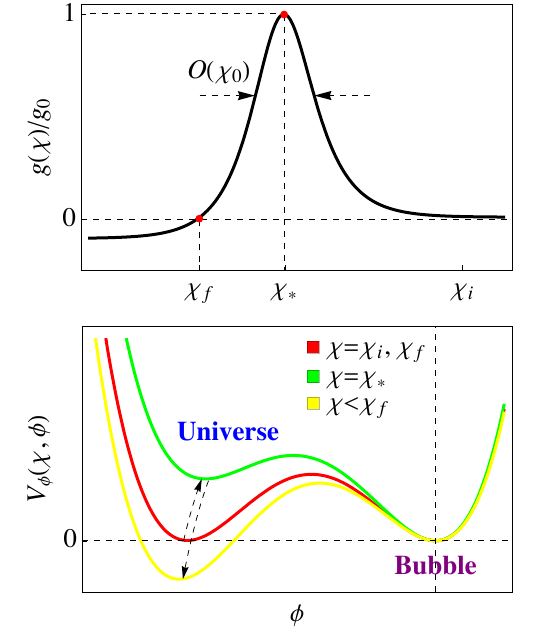}
    \caption{Plots of $g(\chi)/g_0$ (top) and $V_\phi(\chi,\phi)$ (bottom) for \eqref{eq:g(chi)}. During inflation, the value of $g(\chi)$ and thus the energy difference between FV and TV maximize at $\chi=\chi_*$, resulting in the highest tunneling rate. After inflation, $g(\chi<\chi_f)<0$ so that $V_\phi(\chi, \phi_-)<V_\phi(\chi, \phi_+)$, causing the collapse of the bubbles.
    \label{fig:gchi}}
\end{figure}

The first task is to determine the function $g(\chi)$. We set $\phi_-$ and $\phi_+$ as two local minimum of $V_{\phi}(\chi,\phi)$ such that
\begin{eqnarray}\label{eq:phiF}
    \phi_{-}\equiv-\frac{v}{2}\left(1+\sqrt{1-\frac{2g(\chi)}{\lambda}}\right)\,,\quad \phi_{+}\equiv v\,,
\end{eqnarray}
where $\phi_-<\phi_+$. For simplicity, at the beginning of inflation where $\chi=\chi_i$, we set the two vacua degenerate: $V_{\phi}(\phi_-, \chi_i)=V_{\phi}(\phi_+, \chi_i)$, so that the  bubble nucleation is suppressed. This yields a condition $g(\chi_i)=0$. When the inflaton $\chi$ slowly rolls down, $V_\phi(\phi_-, \chi)$ gradually rises and reaches the maximum at $\chi=\chi_*$ and most of the bubbles are nucleated at this point. After this, $\chi$ keeps slowly rolling down along $V_{\chi}$ until the end of inflation where $\chi=\chi_{f}$. During this period, $V_\phi(\phi_-, \chi)$ gradually decreases and finally goes back to $V_{\phi}(\phi_-, \chi_f)=V_{\phi}(\phi_+, \chi_f)$. After inflation, the inflaton decays to the value $\chi<\chi_{f}$ with $V_{\phi}(\phi_-, \chi<\chi_f)<V_{\phi}(\phi_+, \chi<\chi_f)$, 
so the bubbles stop expansion and start to collapse. 
The evolution of $V_\phi(\phi, \chi)$ is illustrated in Fig.~\ref{fig:gchi}.


Therefore, the key feature of $g(\chi)$ is that it should realize the scenario that $V_\phi(\phi_-, \chi)$ monotonically increases for $\chi>\chi_*$ and decreases for $\chi<\chi_*$, together with $V_\phi(\phi_-, \chi_i)=V_\phi(\phi_-, \chi_f)=0$. One simple expression is
\begin{eqnarray}\label{eq:g(chi)}
     g(\chi)=g_0\left[
    \rm{sech}\left(\frac{\chi-\chi_*}{\chi_0}\right)+\alpha~\rm{tanh}\left(\frac{\chi-\chi_*}{\chi_0}\right)\right]\,,
\end{eqnarray}
where $g_0$, $\chi_0$ and $\alpha$ are three positive parameters. The first term in \eqref{eq:g(chi)} creates a peak in $g(\chi)$ at $\chi_*$ with height $g_0>0$ and width $O(\chi_0)$, while the second term allows a zero point $\chi_f$ such that $g(\chi_f)=0$ and a negative $g(\chi)$ when $\chi<\chi_f$ to realize the vacuum exchange, as illustrated in Fig~\ref{fig:gchi}. Since the typical value of $\alpha$
can be sufficiently small%
\footnote{E.g., for $\chi_f=\sqrt{2}M_{\rm Pl}$, $\chi_0= 0.2M_{\rm Pl}$ and $\chi_*=4.66M_{\rm Pl}$, we have $\alpha\sim10^{-6}$.}%
, the $\tanh$ term is safely neglected in the subsequent calculations for simplicity.

\subsection{Inflationary background outside the bubble}
The dynamics of inflation is described by the effective potential
\begin{eqnarray}\label{eq:Veff}
    V_{\rm eff}(\chi)\equiv V(\chi,\phi_{-})=V_\chi(\chi)+V_\phi(\chi, \phi_-)\approx V_{\chi}(\chi)\,.
\end{eqnarray}
In order to satisfy \eqref{eq:phiF} and \eqref{eq:Veff}, we obtain the constraint%
\footnote{The condition for $V_{\rm eff}(\chi)$  to decrease monotonically as $\chi$ decreases (i.e., ${\rm d}V_{\rm eff}(\chi)/{\rm d}\chi>0$) reads  $m^2\chi_0\chi_f>Dg_0 v^4$, where $\chi_0/M_{\rm Pl}\sim O(10^{-1})$, $\chi_f/M_{\rm Pl}\sim O(1)$ and $D\sim O(1)$. This is guaranteed by the first inequality in \eqref{eq:chiDominate} which at least yields $m^2\chi_f^2\gg g_0v^4$.}
\begin{eqnarray}\label{eq:chiDominate}
    0<\frac{2g_0}{\lambda}<1\,,
    \qquad
    V_{\chi}(\chi)\gg V_{\phi}(\chi,\phi_{-})\,.
\end{eqnarray}
Hence, the inflationary background outside the bubble is similar to chaotic inflation%
\footnote{Although the chaotic inflationary model is disfavored by CMB observations, since our main interest is not in inflation itself, here we employ it only for simplicity and illustrative purpose. Our main results remain valid for a broad class of inflationary models.}%
. For definiteness, we choose
\begin{eqnarray}
    m=10^{-5}M_{\rm Pl}\,,
    \quad
    \chi_i=15M_{\rm Pl}\,,
    \quad
    \chi_f=\sqrt{2}M_{\rm Pl}\,,
\end{eqnarray}
so that $V_{\chi}\sim(10^{15}{\rm GeV})^4$ and the total e-fold number $\mathcal{N}\approx60$.
Note that in this model, $\chi$ decreases linearly with time $t$:
\begin{eqnarray}\label{eq:chi(t)}
    \chi(t)\approx\chi_{f}-\sqrt{\frac{2}{3}}mM_{\rm Pl}(t-t_{f})\,.
\end{eqnarray}
While the Hubble parameter $H_{\rm inf}$ in chaotic inflation is also time-dependent, we treat it as a constant $H_{\rm inf}=m\chi_i/(\sqrt{6}M_{\rm Pl})$ for analytical convenience.

\subsection{From tunneling rate to PBH mass function}
With Eq.~\eqref{eq:chi(t)}, one can express the nucleation rate $\Gamma$ in \eqref{eq:f(M)} as a function of PBH mass $M$. 
Conventionally, the FV decay rate can be calculated using semi-classical theory~\cite{Coleman:1977py,Coleman:1980aw}:
\begin{eqnarray}
    \Gamma=A\exp\left[-B(\chi)\right]\,,
    \quad\,
    B(\chi)=S_{\rm E}\left[\phi_{\rm B}\right]-S_{\rm E}\left[\phi_{\rm F}\right]\,,
\end{eqnarray}
where $S_{\rm E}$ is the Euclidean action evaluated with $O(4)$-symmetry, $A$ is a normalization factor and $\phi_{\rm B}$ is the bounce solution of the Euclidean equation of motion (EOM) with appropriate boundary conditions. Although it is difficult to calculate the prefactor $A$~\cite{Ivanov:2022osf,Guada:2020ihz,Wang:2025ooq}, noting that its magnitude is of the order of $H_{\rm inf}^4$~\cite{Linde:1981zj,Linde:1990flp,Weinberg:2012pjx}, we can approximate it as
\begin{eqnarray}\label{gammaH}
    \frac{\Gamma}{H_{\rm inf}^4}\sim \exp\left[-B(\chi)\right]\,.
\end{eqnarray}
Inserting \eqref{gammaH} into \eqref{eq:f(M)}, one obtains
\begin{eqnarray}\label{eq:lnf(M)}
    \ln f(M)=-B(\chi)+\ln C +\frac{1}{2}\ln\left(\frac{\mathcal{M}_{\mathrm{eq}}}{M}\right)\,.
\end{eqnarray}
As the inflaton rolls down, $B$ changes with $\chi$ since the shape of the tunneling potential $V(\phi,\chi)$ varies. Meanwhile, combining \eqref{eq:chi(t)}, \eqref{eq:R(t)} and \eqref{eq:M(R)}, the inflaton field $\chi$ can be expressed as a function of PBH mass $M$ as follows
\begin{eqnarray}\label{eq:chi(M)}
    \chi(M)=\chi_{f}+\frac{mM_{\rm Pl}}{\sqrt{6}H_{\rm inf}}\ln \left(\frac{H_{\rm inf}M}{M_{\rm Pl}^2}\right)\,,
\end{eqnarray}
For example, PBHs with $M_*=10^{20}\,\rm g$ form from bubbles nucleated during inflation when the inflaton rolls to $\chi_*\approx4.66 M_{\rm Pl}$, which is around the last fifth e-folding of inflation. Therefore, \eqref{eq:lnf(M)} and \eqref{eq:chi(M)} imply that once $B(\chi)$ is derived, the mass function $f(M)$ can be expressed in terms of the PBH mass $M$.

\section{Evaluation of the tunneling rate}
\label{sec:tunnelRate}
In this section, we evaluate the exponent $B(\chi)$. We focus on two extremal cases in which it can be formulated analytically. One is the Coleman-De Luccia (CDL) instanton which exists for a special form of potentials such that the curvature scale of the barrier is large compared to the potential energy (in Planck units)~\cite{Coleman:1980aw}. When the curvature scale shrinks, it will gradually approach another case called Hawking-Moss (HM) instanton~\cite{Hawking:1981fz}.


\subsection{Tunneling via CDL instanton}

In CDL case, a fitting function to calculate $B_{\rm CDL}$ exists:
\begin{align}\label{eq:BCDL1}
    B_{\rm CDL}&=\frac{\pi^2(1-\delta)(1+10.07\delta+16.55\delta^2)}{6\lambda(\delta+\omega)^2(1+10\delta)}\nonumber\\
    &\times\left[\frac{1}{\omega}+\frac{11}{2}\frac{\delta}{\omega}+3\left(\frac{\delta}{\omega}\right)^2+\frac{3}{2}\left(\frac{\delta}{\omega}\right)^3\right]\nonumber\\
    &\times\left(1+\sum_{i,j=0}^{4}a_{ij}\delta^i\omega^j\right)\,,
\end{align}
where $\delta$ and $\omega$ are functions of $\chi$ expressed in \eqref{eq:delta(chi)} and below \eqref{eq:BCDL}, respectively, while $a_{ij}$ is a constant matrix defined in \eqref{eq:amatrix}. Inserting \eqref{eq:BCDL1} into \eqref{eq:lnf(M)}, using \eqref{eq:chi(M)} to substitute $\chi$ with $M$, we obtain the mass function of PBHs in CDL case. 

\begin{figure*}[ht]
  \centering
  \includegraphics[width=0.99\linewidth]{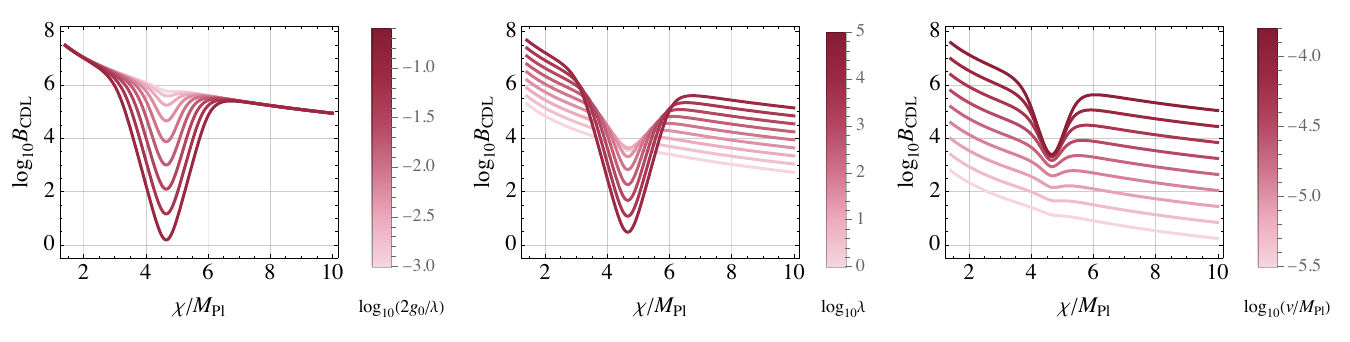}
  \caption{Plot of $B_{\rm CDL}(\chi)$'s dependence on parameters $2g_0/\lambda$, $\lambda$ and $v$, where $\chi_*=4.66M_{\rm Pl}$ is fixed. 
  [left] $\chi_0=0.3M_{\rm Pl}$, $\lambda=10^{4}$ and $v=10^{-4}M_{\rm Pl}$.
  [middle] $\chi_0=0.3M_{\rm Pl}$, $2g_0/\lambda=0.1$ and $v=10^{-4}M_{\rm Pl}$. [right] $\chi_0=0.3M_{\rm Pl}$, $2g_0/\lambda=0.02$ and $\lambda=10^{4}$.
  For a reasonable semi-classical bounce action, we keep $B_{\rm CDL}\gg1$.
  \label{fig:BCDL}}
\end{figure*}

\begin{figure}[ht]
    \includegraphics[width=0.98\linewidth]{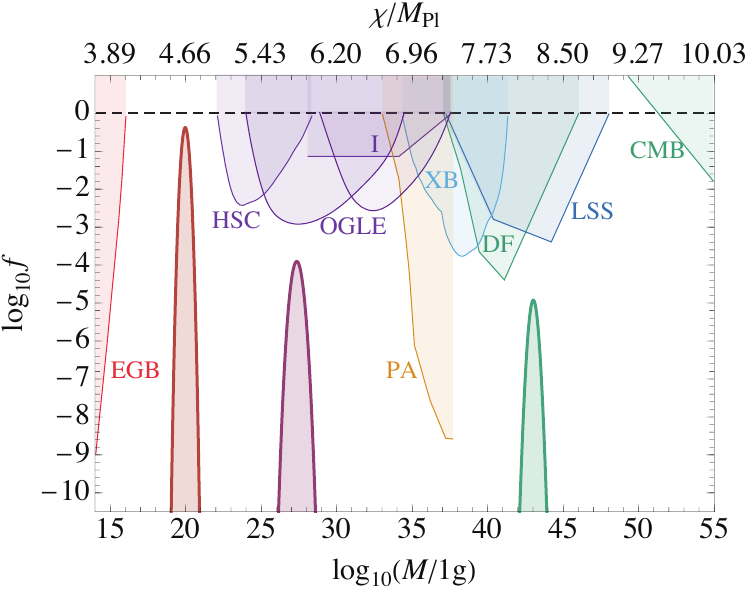}
    \caption{Mass function $f(M)$ of PBHs as DM from collapsing bubbles nucleated via CDL instanton, which stays consistent with current observational constraints \cite{Carr_2021}.
    Evaporation constraints (red) show the extragalactic $\gamma$-ray background (EGB) \cite{Carr:2009jm}. Lensing constraints (purple) come from microlensing by HSC, OGLE and Icarus event (I) \cite{Mroz:2024wia,EROS-2:2006ryy}. Accretion constraints (light blue, yellow) come from X-ray binaries (XB) \cite{Inoue_2017} and CMB anisotropies measured by Planck (PA) \cite{Serpico:2020ehh}. Dynamical constraints (green) are from infalling of halo objects due to dynamical friction (DF) and the CMB dipole (CMB) \cite{Carr:1997cn}. Large-scale structure constraints (dark blue) demonstrate various cosmic structures (LSS) \cite{Carr_2018}.
    The three curves of mass function $f(M)$ from left to right correspond to the choices for model parameters listed in Tab.~\ref{tab:parameter}. The top axis relates the $\chi$ value of inflaton to the PBH mass $M$ through Eq. \eqref{eq:chi(M)}.
   \label{fig:fPBH}}
\end{figure}

In Fig.~\ref{fig:BCDL}, we fix $\chi_*=4.66M_{\rm Pl}$ (which corresponds to $M_*=10^{20}{\rm g}$) and plot $B_{\rm CDL}(\chi)$ for different values of the parameters $g_0$, $\lambda$ and $v$, respectively. 
For CDL instanton to exist and dominate \cite{Weinberg:2005af,Sasaki:1994yj}, the parameters in our model must satisfy the condition given by \eqref{constrainpara}.
It is evident that $B_{\rm CDL}(\chi)$ is minimized at $\chi=\chi_*$. According to \eqref{eq:lnf(M)}, this leads to a sharp peak in the PBH mass function at $M=M_*$. In Fig.~\ref{fig:fPBH}, we choose three sets of parameters (listed in Tab.~\ref{tab:parameter}) and plot the corresponding PBH abundance formed by the vacuum bubbles nucleated via CDL instanton. Depending on the values of parameters, our model can produce PBHs within different mass scales, e.g. PBHs as a dominant candidate of CDM, sub-solar PBHs and supermassive PBHs. Specifically, to realize the peak centered on $M_*=10^{20}\,\rm g$ depicted in Fig.~\ref{fig:fPBH}, parameters should be arranged such that the maximum nucleation rate reaches the order $\exp(-B)\sim10^{-17}$ during inflation.

\begin{table}[ht]
    \centering
    \caption{Parameters sets for the mass function of PBHs in Fig.~\ref{fig:fPBH} from bubbles nucleated via CDL instanton.}
    \label{tab:parameter}
    \begin{tabular}{c|c c c c c}
        Set & $\chi_*/M_{\rm Pl}$ & $2g_0/\lambda$ & $\lambda$ & $v/M_{\rm Pl}$ & $\chi_0/M_{\rm Pl}$ 
        \\
        \hline \hline
         (1) & $4.66$ &$ 0.6$ & $9.33$ & $1.79\times 10^{-4}$ & $0.3$ 
        \\
        \hline
         (2) & $5.8$ & $0.165$ & $10^3$ & $1\times10^{-4}$ & $0.4$ 
        \\
        \hline
         (3) & $8.2$ & $0.99$ & $0.199$ & $1.10\times10^{-3}$ & $1$ 
    \end{tabular}
\end{table}

\subsection{Tunneling via HM instanton}
A HM solution can be found by setting the tunneling field to its local maximum value $\phi=\phi_{\rm top}$ where
\begin{eqnarray}
   \phi_{\rm top}\equiv-\frac{v}{2}\left(1-\sqrt{1-\frac{2g(\chi)}{\lambda}}\right)\,.
\end{eqnarray}
Under the approximation \eqref{eq:chiDominate}, we can approximate $V^{-1}(\phi, \chi)\approx V_\chi^{-1}(1-V_{\phi}/V_\chi)$ so that
\begin{align}\label{eq:BHM}
    B_{\rm HM}&=24\pi^2 M_{\rm Pl}^4\left( \frac{1}{V(\phi_{-},\chi)}- \frac{1}{V(\phi_{\mathrm{top}},\chi)}  \right)\nonumber\\
    &\approx96\lambda\pi^2 v^4\left(\frac{M_{\rm Pl}}{m}\right)^4h(\chi)\,,
\end{align}
where
\begin{eqnarray}\label{eq:hchi}
    h(\chi)\equiv\frac{1}{\chi^4}\left[1-{\frac{2g_0}{\lambda}}{{\rm sech}\left(\frac{\chi-\chi_*} {\chi_0}\right)}\right]^{3/2}\,.
\end{eqnarray}

Under the condition
\begin{eqnarray}\label{eq:HMminimum}
    \left(\frac{\chi_0}{\chi_*}\right)^2\left(\frac{\lambda}{2g_0}-1\right)\lesssim\frac{1}{10}\,,
\end{eqnarray}
a local minimum in $h(\chi)$ appears at $\chi\approx\chi_*$. The general profile of $B_{\rm HM}(\chi)$ is similar to that of $B_{\rm CDL}(\chi)$ in Fig.~\ref{fig:BCDL}. 

We note that in the case where $g_0=0$ so that the non-minimal coupling between inflaton $\chi$ and instanton $\phi$ vanishes, $B_{\rm HM}$ does not stay constant but still increases due to the slow roll of $\chi$. 
The similar behavior happens in CDL case.


\section{Conclusion and discussion}
\label{sec:conclusion}
In this Letter, we propose a two-field model where the instanton field $\phi$ non-minimally couples to the inflaton field $\chi$. By choosing an appropriate coupling function, we realize the scenario where the difference between FV and TV is maximized during inflation when $\chi=\chi_*$, leading to a maximized tunneling rate at this point. Most of the bubbles form at this time. After inflation, the potential in the FV drops below that in the TV. Correspondingly, PBHs form through the collapse of these bubbles, resulting in a sharp peak in the mass function of PBHs.

In particular, we analyze the nucleation of bubbles through CDL and HM instantons in our model, respectively. It is found that the CDL instanton exists within a wide range of parameters. Using a two-dimensional fitting function to evaluate the tunneling probability factor $B_{\rm CDL}$, we find that it is minimized at $\chi=\chi_*$, resulting in a sharp peak in the corresponding PBH mass function. Choosing different values of $\chi_*$, we can realize either PBHs as whole DM, PBHs with astroid mass, or supermassive PBHs. The similar behavior occurs for the bubbles created through HM instantons. We emphasize that even a slight variation in the tunneling potential can significantly affect the tunneling rate and, consequently, the resulting PBH abundance. We note that since this scenario does not require the enhancement of primordial curvature perturbations, it does not necessarily produce significant Induced Gravitational Wave (IGW) signals. Hence, even if IGW signal is not observed in the future GW detections, the existence of PBHs cannot be ruled out.

\section*{Acknowledgments}
We thank Misao Sasaki, Jaume Garriga, Cristiano Germani, Kazunori Kohri, Przemek Mróz and Jun'ichi Yokoyama for useful discussions and valuable comments. 
Y.Z. and H.W. are supported by  the Fundamental Research Funds for the Central Universities, and by the Project 12475060 and 12047503 supported by NSFC, Project 24ZR1472400 sponsored by Natural Science Foundation of Shanghai, and Shanghai Pujiang Program 24PJA134. T.S. gratefully acknowledges support from JSPS KAKENHI grant (Grant Number JP23K03411).

\bibliographystyle{apsrev4-1}
%


\appendix

\onecolumngrid

\section{Calculation for CDL instanton}\label{app:CDL}
An analytic bounce action for CDL instanton can only be acquired in thin-wall approximation \cite{parke1983gravity}. However, for a general quartic potential
\begin{eqnarray}\label{generalV}
    V(\phi)=\lambda\phi^4-\mu\phi^3+\frac{1}{2}\nu^2\phi^2+V_0\,,
\end{eqnarray}
Ref. \cite{Sasaki:1994yj} proposed a 2D polynomial fitting function for $B_{\rm CDL}$. Defining the following parameters from the coefficients of the potential \eqref{generalV}
\begin{eqnarray}\label{eq:H0<}
    \delta\equiv 1-\frac{2\lambda\nu^2}{\mu^2}\in(0,1)\,,
    \qquad
    H_0\equiv \frac{1}{\nu}\left(\frac{V_0}{3M_{\rm Pl}^2}\right)^{1/2}\in \left(0,\frac{1}{2}\right)\,,
\end{eqnarray}
the constraint for $\delta$ which characterizes the shape of the quartic polynomial ($\delta\to0$ for two degenerate vacua and $\delta\to1$ for a vanishing barrier) is to ensure that the potential contains FV and TV, while the constraint for $H_0$ is from the condition for CDL instanton to dominate the tunneling process in dS: 
\begin{align}
\frac{V_0}{3M_{\rm Pl}^2}< \frac{|V''(\phi_{\rm top})|}{4}\,.
\end{align}

Then $B_{\rm CDL}$ can be expressed in terms of $H_0$ and $\delta$ as a two-dimensional fitting function:%
\begin{eqnarray}\label{eq:BCDL}
    B_{\rm CDL}=\frac{\pi^2}{12}\frac{1-\delta}{2\lambda}
    \frac{4(1+10.07\delta+16.55\delta^2)}{\omega(\delta+\omega)^2(1+10\delta)}
    \times\left[1+\frac{11}{2}\delta+3\frac{\delta^2}{\omega}+\frac{3}{2}\frac{\delta^3}{\omega^2}\right]
    \times\left(1+\sum_{i,j=0}^{4}a_{ij}\delta^i\omega^j\right)\,,
\end{eqnarray}
where $\omega\equiv\sqrt{\delta^2+H_0^2}$ and the coefficient $a_{ij}$ reads
\begin{eqnarray}\label{eq:amatrix}
     a_{ij}=\begin{pmatrix}
0 & -0.1617 & 0 & -5.507 & 0 \\
0 & -11.34 & 30.17 & 0 & -21.69 \\
12.09 & -33.24 & 0 & 41.29 & 0 \\
7.728 & 0 & -19.34 & 0 & 0 \\
0 & 0 & 0 & 0 & 0 \\
\end{pmatrix}\,.
\end{eqnarray}

Now we compare \eqref{generalV} with our potential \eqref{eq:Vphi} to express $\delta$ and $H_0$, which results in the following correspondence:
\begin{eqnarray}\label{replacement}
    \mu\to 2\lambda v\left(1+\sqrt{\xi}\right)\,,
    \qquad
    \nu\to \sqrt{\lambda \left(3\sqrt{\xi}+\xi\right)}\,,
    \qquad
    V_0\to V_\chi(\chi)\,,
\end{eqnarray}
where
\begin{eqnarray}
\xi\equiv1-\frac{2g(\chi)}{\lambda}\in(0,1)\,.
\end{eqnarray}
We note that in order to approach the form of \eqref{generalV},  we have shifted potential \eqref{eq:Vphi} along the $\phi$-direction such that FV is located at $\phi_-=0$ and the term proportional to $\phi$ vanishes. Since we are interested in calculating the tunneling rate at every point of $\chi$, we fix the value of $\chi$ so that $g(\chi)$ is also a fixed value. Then inserting \eqref{replacement} into \eqref{eq:H0<}, $\delta$ and $H_0$ can be expressed in terms of parameters and $\xi$ in our model such that
\begin{eqnarray}\label{eq:delta(chi)}
    \delta=\frac{1-\sqrt{\xi}}{1+\xi+2\sqrt{\xi}}\,,
    \qquad
    H_0=\frac{m\chi}{2\sqrt{3}M_{\rm Pl} v}\frac{1}{\sqrt{\lambda\left(\xi+3\sqrt{\xi}\right)}}\,.
\end{eqnarray}
After substituting \eqref{eq:delta(chi)} into \eqref{eq:BCDL} it is straightforward to obtain $B_{\rm CDL}(\chi)$. 

We note that \eqref{eq:H0<} together with \eqref{eq:chiDominate} yields a constraint in the $\lambda-v$ parameter space%
:

\begin{eqnarray}\label{constrainpara}
    v^2f(\xi)
     \ll\frac{m^2\chi^2}{\lambda v^2}<3M_{\rm Pl}^2\left(\xi+3\sqrt{\xi}\right)\,,     
\end{eqnarray}
where 
\begin{eqnarray}
f(\xi)\equiv\left(1-\sqrt{\xi}\right)+\frac{1}{2}(1-\xi)\left(5+2\sqrt{\xi}\right)-\frac{(1-\xi)^2}{8}\,,
\end{eqnarray}
monotonically decreases from roughly $3.5$ to $0$ when $\xi$ increases from $0$ to $1$. Since $\xi$ varies from $1$ to $\xi_*=1-2g_0/\lambda$ and then returns to $1$ during inflation, the sufficient conditions for \eqref{constrainpara} can be further written as
\begin{eqnarray}
    v^2f(\xi_*)
     \ll\frac{m^2\chi_f^2}{\lambda v^2}\,,
     \qquad
     \frac{m^2\chi_i^2}{\lambda v^2}<3M_{\rm Pl}^2\left(\xi_*+3\sqrt{\xi_*}\right)\,.
\end{eqnarray}


\end{document}